\documentclass[aps,prl,twocolumn,superscriptaddress,showpacs]{revtex4}

\usepackage{epsf,times}
\newcommand{\cutText}{1}  % if this is set to 1, cut text to remove 63
\def\np#1#2#3   {{ Nucl. Phys.} {\bf#1}, #2 (#3). }
\def\pcps#1#2#3 {{ Proc. Cam. Phil. Soc.} {\bf#1}, #2 (#3). }
\def\pl#1#2#3   {{ Phys. Lett.} {\bf#1}, #2 (#3). }
\def\plc#1#2#3   {{ Phys. Lett.} {\bf#1}, #2 (#3); }
\def\prep#1#2#3 {{ Phys. Rep.} {\bf#1}, #2 (#3). }
\def\prev#1#2#3 {{ Phys. Rev.} {\bf#1}, #2 (#3). }
\def\prl#1#2#3  {{ Phys. Rev. Lett.} {\bf#1}, #2 (#3). }
\def\prs#1#2#3  {{ Proc. Roy. Soc.} {\bf#1}, #2 (#3). }
\def\ptp#1#2#3  {{ Prog. Th. Phys.} {\bf#1}, #2 (#3). }
\def\rmp#1#2#3  {{ Rev. Mod. Phys.} {\bf#1}, #2 (#3). }
\def\rpp#1#2#3  {{ Rep. Prog. Phys.} {\bf#1}, #2 (#3). }
\def\zp#1#2#3   {{ Zeit. Phys.} {\bf#1}, #2 (#3). }
\def\epj#1#2#3   {{ Eur. Phys. Jour.} {\bf#1}, #2 (#3). }
\def\nim#1#2#3   {{ Nucl. Instr. Meth.} {\bf#1}, #2 (#3). }
\def \D0  {${\rm D{\not\!{\rm O}}}$}

\newcommand{\met}{\mbox{$\protect \raisebox{.3ex}{$\not$}E_T$}}

\begin{document}
\title{A Search for $t \rightarrow \tau\nu q$ in $t\bar{t}$ Production}

%Authors
\affiliation{Institute of Physics, Academia Sinica, Taipei, Taiwan 11529, Republic of China} 
\affiliation{Argonne National Laboratory, Argonne, Illinois 60439} 
\affiliation{Institut de Fisica d'Altes Energies, Universitat Autonoma de Barcelona, E-08193, Bellaterra (Barcelona), Spain} 
\affiliation{Baylor University, Waco, Texas  76798} 
\affiliation{Istituto Nazionale di Fisica Nucleare, University of Bologna, I-40127 Bologna, Italy} 
\affiliation{Brandeis University, Waltham, Massachusetts 02254} 
\affiliation{University of California, Davis, Davis, California  95616} 
\affiliation{University of California, Los Angeles, Los Angeles, California  90024} 
\affiliation{University of California, San Diego, La Jolla, California  92093} 
\affiliation{University of California, Santa Barbara, Santa Barbara, California 93106} 
\affiliation{Instituto de Fisica de Cantabria, CSIC-University of Cantabria, 39005 Santander, Spain} 
\affiliation{Carnegie Mellon University, Pittsburgh, PA  15213} 
\affiliation{Enrico Fermi Institute, University of Chicago, Chicago, Illinois 60637} 
\affiliation{Joint Institute for Nuclear Research, RU-141980 Dubna, Russia} 
\affiliation{Duke University, Durham, North Carolina  27708} 
\affiliation{Fermi National Accelerator Laboratory, Batavia, Illinois 60510} 
\affiliation{University of Florida, Gainesville, Florida  32611} 
\affiliation{Laboratori Nazionali di Frascati, Istituto Nazionale di Fisica Nucleare, I-00044 Frascati, Italy} 
\affiliation{University of Geneva, CH-1211 Geneva 4, Switzerland} 
\affiliation{Glasgow University, Glasgow G12 8QQ, United Kingdom} 
\affiliation{Harvard University, Cambridge, Massachusetts 02138} 
\affiliation{Division of High Energy Physics, Department of Physics, University of Helsinki and Helsinki Institute of Physics, FIN-00014, Helsinki, Finland} 
\affiliation{University of Illinois, Urbana, Illinois 61801} 
\affiliation{The Johns Hopkins University, Baltimore, Maryland 21218} 
\affiliation{Institut f\"{u}r Experimentelle Kernphysik, Universit\"{a}t Karlsruhe, 76128 Karlsruhe, Germany} 
\affiliation{High Energy Accelerator Research Organization (KEK), Tsukuba, Ibaraki 305, Japan} 
\affiliation{Center for High Energy Physics: Kyungpook National University, Taegu 702-701; Seoul National University, Seoul 151-742; and SungKyunKwan University, Suwon 440-746; Korea} 
\affiliation{Ernest Orlando Lawrence Berkeley National Laboratory, Berkeley, California 94720} 
\affiliation{University of Liverpool, Liverpool L69 7ZE, United Kingdom} 
\affiliation{University College London, London WC1E 6BT, United Kingdom} 
\affiliation{Massachusetts Institute of Technology, Cambridge, Massachusetts  02139} 
\affiliation{Institute of Particle Physics: McGill University, Montr\'{e}al, Canada H3A~2T8; and University of Toronto, Toronto, Canada M5S~1A7} 
\affiliation{University of Michigan, Ann Arbor, Michigan 48109} 
\affiliation{Michigan State University, East Lansing, Michigan  48824} 
\affiliation{Institution for Theoretical and Experimental Physics, ITEP, Moscow 117259, Russia} 
\affiliation{University of New Mexico, Albuquerque, New Mexico 87131} 
\affiliation{Northwestern University, Evanston, Illinois  60208} 
\affiliation{The Ohio State University, Columbus, Ohio  43210} 
\affiliation{Okayama University, Okayama 700-8530, Japan} 
\affiliation{Osaka City University, Osaka 588, Japan} 
\affiliation{University of Oxford, Oxford OX1 3RH, United Kingdom} 
\affiliation{University of Padova, Istituto Nazionale di Fisica Nucleare, Sezione di Padova-Trento, I-35131 Padova, Italy} 
\affiliation{University of Pennsylvania, Philadelphia, Pennsylvania 19104} 
\affiliation{Istituto Nazionale di Fisica Nucleare Pisa, Universities of Pisa, Siena and Scuola Normale Superiore, I-56127 Pisa, Italy} 
\affiliation{University of Pittsburgh, Pittsburgh, Pennsylvania 15260} 
\affiliation{Purdue University, West Lafayette, Indiana 47907} 
\affiliation{University of Rochester, Rochester, New York 14627} 
\affiliation{The Rockefeller University, New York, New York 10021} 
\affiliation{Istituto Nazionale di Fisica Nucleare, Sezione di Roma 1, University of Rome ``La Sapienza," I-00185 Roma, Italy} 
\affiliation{Rutgers University, Piscataway, New Jersey 08855} 
\affiliation{Texas A\&M University, College Station, Texas 77843} 
\affiliation{Istituto Nazionale di Fisica Nucleare, University of Trieste/\ Udine, Italy} 
\affiliation{University of Tsukuba, Tsukuba, Ibaraki 305, Japan} 
\affiliation{Tufts University, Medford, Massachusetts 02155} 
\affiliation{Waseda University, Tokyo 169, Japan} 
\affiliation{Wayne State University, Detroit, Michigan  48201} 
\affiliation{University of Wisconsin, Madison, Wisconsin 53706} 
\affiliation{Yale University, New Haven, Connecticut 06520} 
\author{A.~Abulencia}
\affiliation{University of Illinois, Urbana, Illinois 61801}
\author{D.~Acosta}
\affiliation{University of Florida, Gainesville, Florida  32611}
\author{J.~Adelman}
\affiliation{Enrico Fermi Institute, University of Chicago, Chicago, Illinois 60637}
\author{T.~Affolder}
\affiliation{University of California, Santa Barbara, Santa Barbara, California 93106}
\author{T.~Akimoto}
\affiliation{University of Tsukuba, Tsukuba, Ibaraki 305, Japan}
\author{M.G.~Albrow}
\affiliation{Fermi National Accelerator Laboratory, Batavia, Illinois 60510}
\author{D.~Ambrose}
\affiliation{Fermi National Accelerator Laboratory, Batavia, Illinois 60510}
\author{S.~Amerio}
\affiliation{University of Padova, Istituto Nazionale di Fisica Nucleare, Sezione di Padova-Trento, I-35131 Padova, Italy}
\author{D.~Amidei}
\affiliation{University of Michigan, Ann Arbor, Michigan 48109}
\author{A.~Anastassov}
\affiliation{Rutgers University, Piscataway, New Jersey 08855}
% ...
\collaboration{CDF Collaboration}
\noaffiliation

\date{\today}

\begin{abstract}

We present a search for $t\bar{t}$ events with a
tau lepton in the final state.  The data sample corresponds
to an integrated luminosity of $194$~pb$^{-1}$ collected with
the CDF II detector from $p\bar{p}$ collisions
at a center of mass energy of 1.96~TeV.  We observe two events with 
an expected signal of $1.0\pm0.2$ events and
a background of $1.3\pm0.3$ events.  We 
determine a 95\% confidence level upper limit on 
$r_{\tau}$, 
the ratio of the measured rate of $t \rightarrow \tau\nu q$ 
to the expectation, of $5.2$.

\end{abstract} 

\maketitle

%
%*********************************************
%%%%%%%%%%%%%%%%%%% ARTICLE TEXT %%%%%%%%%%%%%%%%%%%%
%************************************

An experimental investigation of the interactions among the massive
fermions of the third generation - the top (t) and bottom (b) quarks,
the tau ($\tau$) lepton and tau neutrino ($\nu_\tau$) - has the potential to yield powerful insights
into the puzzles of flavor and fermion mass.  There is no adequate explanation
for the comparatively large masses of the third generation
particles~\cite{PDG}, and we do not understand why there appear
to be three and only three generations.
A significant deviation in the number of observed
$t \to \tau \nu q$\ candidates from the rate predicted by the standard
model
could indicate an anomalous coupling among the third generation particles.
Extensions of the standard model 
%that have particles lighter than the top quark
could lead to alternative modes of top quark decay that enhance
the top branching fraction to this final state. One example is the
minimal supersymmetric standard model~\cite{higgs,susyHiggs-top,higgs2},
where the top quark could decay into a $b$-quark and a charged Higgs boson
with subsequent decay into a tau lepton and tau neutrino.  Other
possibilities include $R$-parity violating SUSY decays of
top~\cite{rpv-top} and new $Z^\prime$
bosons with non-universal couplings~\cite{nu-zprime}.

In this letter, we search for $t \to \tau \nu q$\ decays
in 194 $\pm 11$ pb$^{-1}$ of $p\bar{p}$\ collisions
collected by the CDF collaboration at the $\sqrt{s}$ =1.96~TeV Tevatron
collider at Fermilab.  The top quark, discovered at Fermilab by
the CDF and {${\rm D{\not\!{\rm O}}}$} experiments in 1995~\cite{cdf-top-disc,d0-top-disc}, is
predominantly produced via pair production,
where the next to leading order quantum chromodynamics (QCD) theoretical 
prediction~\cite{ttbar-xsect,ttbar-xsect2}
for the cross section 
%has a precision of below 15\%.  
is $6.7^{+0.7}_{-0.9}$~pb~.  In
the standard model, the top quark is expected to decay with a branching
fraction of almost 100\% into a $W$ boson and a $b$-quark.
%The branching fraction for the decay of a $W$ boson to $\tau \nu$\ is
%measured to be 10.74 $\pm$ 0.27 \%~\cite{PDG}.
The world average of the measurements of the branching fraction for the decay 
of a $W$ boson to $\tau \nu$\ is 10.74 $\pm$ 0.27 \%~\cite{PDG}.
%The only previous search for $t \to \tau \nu q$~\cite{r1tau} can be
%interpreted to exclude
%anomalous rates above 18 times that expected in the standard model at
%95\% confidence level.}
Using the procedure described in this letter we interpret the results
of the only previous search for $t \to \tau \nu q$ in $t\bar{t}$ 
production~\cite{r1tau} to exclude
anomalous rates above 18 times that expected in the standard model at
95\% confidence level.

In this search, we select $t\bar{t}$\ candidates where one top decays to $\tau \nu b$,
and identify the tau lepton by its semi-hadronic decay, which accounts for 
approximately 65\% of tau decays~\cite{PDG}.
We do not search for tau lepton decays to electrons and muons,
as these are difficult to distinguish from electrons and muons
directly from $W$ boson decay.  We require the other top to
decay to either $e \nu b$ or to $\mu \nu b$
in order to utilize the efficient high $p_{T}$~\cite{var} \ electron/muon 
triggers
and to reduce the background from
%QCD
 multi-jet production.
We require significant missing transverse energy from the neutrinos,
and at least two jets with high $E_{T}$ ~\cite{var}, though
we make no requirement on the heavy flavor content of the jets.
Finally, we apply several novel kinematic and topological requirements
designed to reject specific backgrounds.

{\ifnum\cutText=0
The data used for this analysis were collected by the CDF Collaboration
from March 2002 to September 2003.
\fi} 
The CDF II detector~\cite{det} is an azimuthally and
forward-backward symmetric apparatus built to study the physics of
$p\bar{p}$ collisions at 
$\sqrt{s}$ of  $1.96$~TeV.  
The detector contains a
charged-particle tracking system inside a $1.4$~T field generated by a
solenoid coaxial
with the $p$ and $\bar{p}$ beams.  A silicon microstrip detector
provides track measurements between $1.5$ and $28$~cm in radius from
the beam axis for charged particles with pseudorapidity, $|\eta|<2$~\cite{var}.
A $3.1$~m long open cell drift chamber measures track position
at $96$ 
points at radii between $40$ and $137$~cm for particles with
$|\eta|<1$.

Segmented electromagnetic and
hadronic sampling calorimeters surround the tracking volume and cover
the range $|\eta|<3.6$.  
The central
($|\eta|\stackrel{<}{\sim}1$) calorimeters, in which
$\tau$ decays are identified in this analysis, are divided into
towers with segmentation in azimuthal angle of $15$ degrees and
in pseudorapidity of about $0.1$.  The 
central electron shower detector (CES)
consists of proportional chambers with
wires and cathode strips arranged orthogonally
with pitch varying from $1.4$ to
$2.0$~cm located at a depth of 6 radiation lengths within the
electromagnetic calorimeter, 
at the position where the lateral profile of the shower is maximum.  
This fine segmentation of the CES
measures electromagnetic shower position with $\approx 3$~mm
resolution and allows reconstruction of the
boosted $\pi^0\to\gamma\gamma$ produced in tau decays.  
A set of drift chambers located outside the
hadron calorimeters and a second set outside a $60$~cm iron shield
detect muon candidates with $|\eta|<0.6$.  Additional chambers and
scintillator counters extend this muon coverage in 
$0.6<|\eta|<1.0$ for most azimuthal angles.
The luminosity is determined to an accuracy of 6\% using gas  
Cherenkov counters covering $3.7<|\eta|<4.7$ which measure the average number
of inelastic $p\bar{p}$ collisions per bunch crossing.

The online triggers used in this analysis 
select samples with at least one high 
$p_T$ 
central electron or muon
candidate.  The electron trigger requires candidates to have
$E_T$ greater than $18$~GeV,
and the muon trigger requires a candidate with
$p_T>18$~GeV/$c$~\cite{ttbar-kinem-prd}.  
Further identification requirements are placed to select a pure sample
of electrons and muons and are described in detail 
elsewhere~\cite{wz-prod-prl,dilep-prl}.
Neutrinos escape the
calorimeter undetected and result in missing transverse energy
($\met$)
which is measured by balancing the calorimeter energy in the transverse 
plane.
We require $\met>20$~GeV after corrections for identified
muons.

Semi-hadronic decays of taus produced in $W$ decay have a distinctive
signature of narrow, isolated jets with low charged track
multiplicity.  The
calorimeter measures the visible energy of the tau jet, while the
central
tracker and CES determine the narrowness and multiplicity.
A tau candidate requires a tau calorimeter cluster and a central track with a 
minimum $p_T$ of $4.5$~GeV/$c$ pointing
to
the cluster.
A tau calorimeter cluster requires a tower with $E_T\ge6$~GeV and no
more than five adjacent towers with $E_T>1$~GeV.
  
After 
lepton
candidate selection, we impose additional requirements 
on the isolation of the tau lepton
to reduce
backgrounds from jets.
A cone 
is formed around the seed track with a variable angular
radius,
$\theta_{\rm\scriptstyle cone}={\rm Min}\{0.17,(5~{\rm GeV})/E_{\rm\scriptstyle tau~cluster}\}$~rad.
The tau candidate is required to have one or three tracks in the signal cone
to be consistent with the dominant decay modes of the tau.  If there are three
tracks the sum of the electric charges must be equal to $\pm 1$.
Candidate $\pi^0$s
are identified in the calorimeter from clusters of energy
observed in the CES.  The
tau $p_T$ is estimated to be the sum of the 
seed
track $p_T$ plus the sum
of the $\pi^0$ $E_T/c$.  The tau $p_T$ is required to be 
greater than
$15$~GeV/$c$, and the invariant mass of the $\pi^0$s and the tracks is
required to be 
less than
$1.8$~GeV to be consistent with 
the mass of the tau.
An isolation annulus is
defined around the tau cone extending from the 
cone edge, $\theta_{\rm\scriptstyle cone}$, to
0.52 radians in which no tracks or $\pi^0$ candidates may be present.  
Calorimeter towers within the isolation annulus
are required to have $E_T$ less than 6\% of the tau 
$E_T$.
Additional requirements are imposed to
reject tau candidates that resemble electrons or muons based on track
and calorimeter characteristics.  
To remove electrons, the energy in the hadronic calorimeter divided
 by the track momentum sum of the tau candidate, $E_{had}/\sum p$, is required 
to be greater than $0.15$.  To remove muons,
 the $E_T$ of the calorimeter cluster energy
 associated with the tau candidate divided by the $p_T$ of the seed track,
$E_T/p_T$, is required to be greater than $0.5$.  The combined tau 
identification and isolation efficiency is 35\%.

{\ifnum\cutText=0
In addition to the electron or muon, the tau candidate and the missing
transverse energy, we require at least two jets, corresponding to
the expected number of $b$
quarks produced in the $t\bar{t}$ decays.
We require that these jets have $|\eta| <$ 2 and that
the first and second highest jets have $E_T$ 
greater than $25$~GeV and $15$~GeV, respectively.  
\else
We also require at least two jets with $|\eta| <$ 2 and that
the first and second highest $E_T$ jets have $E_T$ 
greater than $25$~GeV and $15$~GeV, respectively.  
\fi}   
The event $H_T$, defined as 
the scalar sum of the electron $E_T$ or muon $p_T$, the tau $p_T$, the
$\met$ and the $E_T$ of the jets, must exceed $205$~GeV.  The 
$H_T$ and lead jet $E_T$ requirements were chosen by a two-dimensional 
optimization that maximized the likelihood ratio of signal plus background
to background-only.  These requirements reduce the backgrounds from 
$W$ bosons produced
in association with jets by $\approx 85\%$, while removing
only $\approx 5\%$ of the $t\bar{t}$ signal.

{\ifnum\cutText=0
\begin{figure}[tbp]
\centerline{\epsfxsize 3.5 truein \epsfbox{BP_wtaunu_trackmult_edited4.eps}}   
 \caption{The number of tracks in the reconstructed tau in
 $W\to\tau\nu$ candidates}
 \label{fig:ntrack}
\end{figure}
\fi} 

We determine the efficiency of the selection cuts by simulating standard 
model $t\bar{t}$ detection with the PYTHIA~\cite{pythia} event generator,
the TAUOLA~\cite{tauola} tau decay simulation and a GEANT-based
%model~\cite{GEANT4} of the CDF detector and trigger response.
model~\cite{GEANT4} of the CDF detector.
%{\ifnum\cutText=0
We independently determine the electron, muon and tau
identification 
and trigger 
efficiencies from the data.
%\fi} 
Electron and muon efficiencies are determined using $Z\to e^+e^-$
and $Z\to\mu^+\mu^-$ events respectively with one of the electrons (muons) 
required to pass tight identification cuts and the other electron (muon) 
used to determine the efficiency. 
The tau identification efficiency is determined by comparing numbers of
observed $W\to\tau\nu$ to the prediction of the simulation.
%Where there are differences
%between the data and the simulation, 
%we derive correction factors 
%($<5\%$ in most cases) which are applied to results of the simulation.
Correction factors, defined as the ratio of the data and MC
efficiencies, are applied to the simulation results. In most
cases correction factors are within 5\% of 1.
{\ifnum\cutText=0
Figure~\ref{fig:ntrack} compares the distribution of
the number of tracks in the tau cone for $W\to\tau\nu$ candidates
to the simulation with the addition of expected 
backgrounds from $W\to
e\nu$, determined from the data, and $Z\to\tau\tau$, determined
from Monte Carlo.
\fi} 
In the case of tau identification,
we find that the ratio of the 
efficiency in data to that in simulation is
$0.90\pm0.06$.  
{\ifnum\cutText=0
In addition to
contributions from the uncertainty in lepton identification
efficiency, there are significant uncertainties from the modeling of
the jet energy response in the calorimeter and in the simulation of
initial and final state QCD radiation.
\fi} 
Table~\ref{tab:sys} summarizes the systematic uncertainties in the
efficiency for detecting the signal process.  

\begin{table}[tbp]
\begin{center}
{\ifnum\cutText=0
{\small
\begin{tabular}{lc}
\hline
\hline
 Source  & signal prediction uncertainty \\
\hline
Jet Energy Scale & $\pm$6$\%$  \\
Electron and Muon Identification &  $\pm$5$\%$  \\
Tau Identification &  $\pm$6$\%$  \\
Top Production Model &  $\pm$7$\%$  \\
Initial State Radiation &  $\pm$7$\%$  \\
Final State Radiation &  $\pm$7$\%$  \\
Parton Distribution Functions &  $\pm$1$\%$  \\
\hline 
Total &  $\pm$16$\%$  \\
\hline
\hline
%Events measured in Data & 2 \\
\end{tabular}}
\else
{\small
\begin{tabular}{lc}
\hline
\hline
 Source  & signal prediction uncertainty \\
\hline
Jet Energy Scale & $\pm$6$\%$  \\
Electron and Muon Identification &  $\pm$5$\%$  \\
Tau Identification &  $\pm$6$\%$  \\
Top Production Model &  $\pm$7$\%$  \\
Parton Distribution Functions &  $\pm$1$\%$  \\
Quark and Gluon Radiation Model &  $\pm$10$\%$  \\
\hline 
Total &  $\pm$16$\%$  \\
\hline
\hline
%Events measured in Data & 2 \\
\end{tabular}}
\fi} 
\end{center}
\caption{Summary of systematic uncertainties in the identification of
$t\bar{t}$ (signal) events}
\label{tab:sys}
\end{table}

Our dominant background is $W$ bosons produced in association with
jets, where the $W$ decays to $e\nu$ or $\mu\nu$ and 
one of the jets is misidentified as a tau.
We determine the number of such events we expect 
from the data.  We first find a sample of 
relaxed
tau candidates passing all 
tau identification
requirements
except the isolation, mass and track quality requirements.  This sample
is dominated by jets rather than tau leptons.  To that
sample, we apply a 
%relative 
fake rate, which is the probability that
those 
relaxed
tau candidates will pass all identification requirements.  
%{\ifnum\cutText=0
The
%relative 
fake rate of jets to identified tau candidates 
is measured in four independent jet
data samples; 
three of the samples were selected with different jet $E_T$ 
threshold requirements and the fourth sample was selected using a requirement
on the sum of the $E_T$ of 
all calorimeter towers in the event.
%\else
%The
%relative fake rate of jets to identified tau candidates 
%is measured in four independent jet
%data samples.
%\fi} 
The 
%relative 
fake rate is
parameterized as a function of jet $E_T$ and
isolation of the jet in the calorimeter.  The full spread in the measured fake
rates of the four different samples is 26\% of the fake rate, which we take as our
estimate of the systematic uncertainty in this procedure.

Events with electrons or muons that fake tau candidates are another significant
background source.  These events originate primarily from the production of $Z$
bosons, decaying to $e^+e^-$ or $\mu^+\mu^-$, in association with
extra jets.  
In the case of electrons, the event can be a
background when the electron energy is poorly reconstructed; in the
case of muons, a muon can fake a tau if the muon suffers a
catastrophic energy loss in the calorimeter.
To estimate the electron background, we 
first
measure an electron to tau fake rate from the data using $Z\rightarrow e^+e^-$ events.
We then scale the number of events with tau candidates that fail the electron
rejection requirement but pass all other analysis requirements
by the electron to tau fake rate.
An  ALPGEN~\cite{alpgen}     
interfaced with HERWIG~\cite{herwig} 
simulation of 
$Z\rightarrow\mu^+\mu^-$ events with extra jets 
is used to predict the
muon background, and we confirm 
the
modeling of the muon
energy response in the calorimeter using $Z\to\mu^+\mu^-$ events in
the data.

\begin{figure}[tbp]
\epsfxsize=3.5 truein\epsfbox{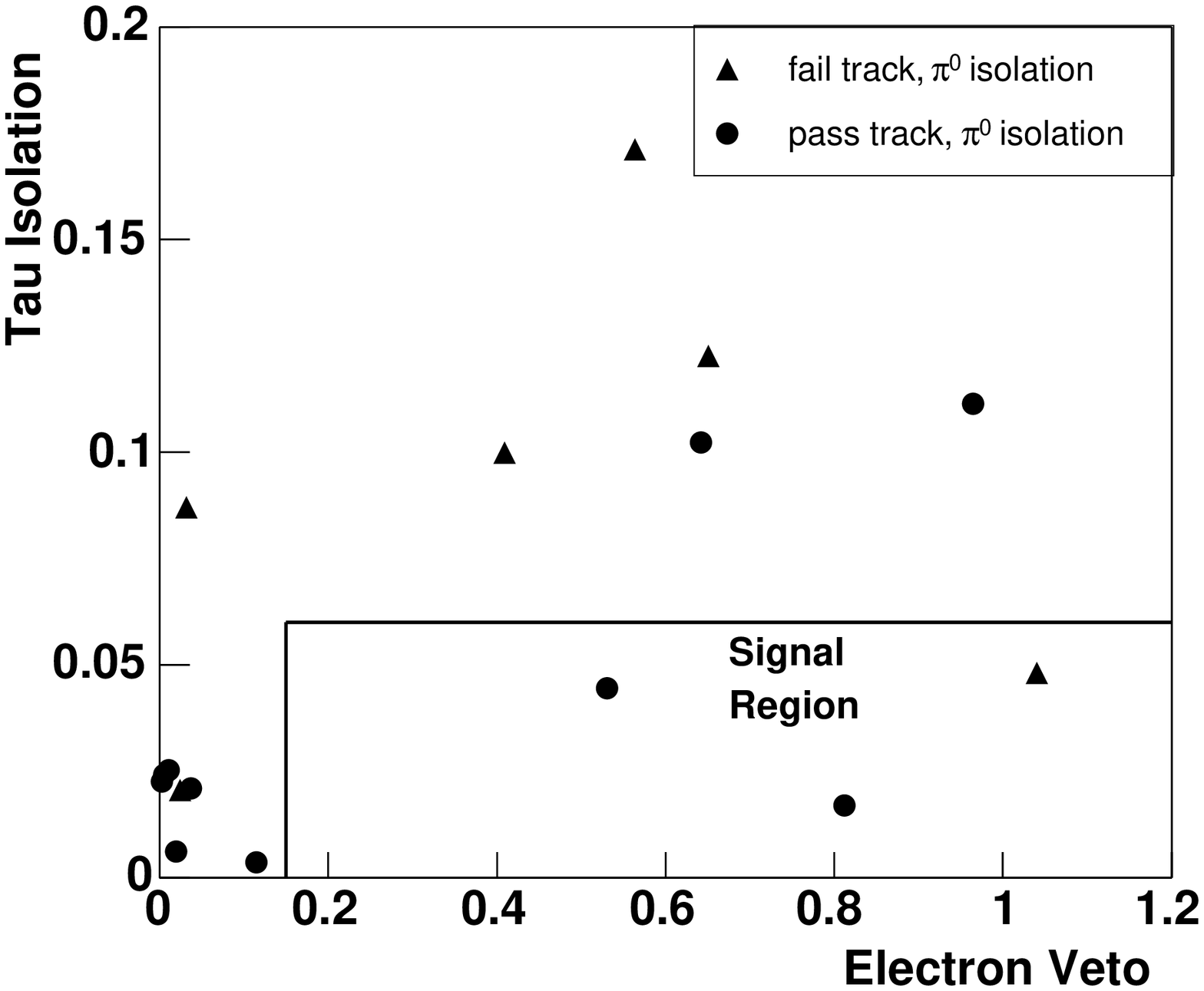}
\caption{Evidence for fake tau background.  For events
 with a tau candidate, an electron, $\met$ and two jets,
 we show the ratio of energy in the tau calorimeter isolation cone divided
 by tau $p_T$ (``tau isolation'') vs.\ the ``electron veto''
 variable, the energy in the hadronic calorimeter divided
 by the track momentum sum of the tau candidate, $E_{had}/\sum p$.
 Events with fake tau candidates from jets are at high tau isolation;
 di-electron events appear at low isolation and low $E_{had}/\sum p$.
 The analysis requirements are shown by lines on the plots.  
The triangular markers identify events which fail the tau identification 
requirements of no $\pi^0$s and tracks in the isolation annulus.} 
\label{fig:eleBgnd}
\end{figure}

Figure~\ref{fig:eleBgnd} shows evidence of fake tau background from
jets and electrons in the electron + tau candidate data with relaxed tau
identification requirements.  
The tau contribution to the region shown outside of the signal
requirements is expected to be a small fraction of that inside.

Another class of background events results from processes other than
$t\bar{t}$ production that create
tau leptons in association with electrons or muons.
The largest of these backgrounds is from $Z$ boson production in association
with jets where $Z\rightarrow \tau^+\tau^-$ with one fully leptonic
and one semi-hadronic tau decay.
In this process, 
the energy spectra of the leptons, jets and the $\met$ are softer than
the predictions from $t\bar{t}$ production.  As a result, 
our $H_T$ requirement reduces this background by about
40\%.  However, even with such a requirement the previous search for this
decay chain at CDF~\cite{r1tau} predicted a higher number of
$Z\to\tau^+\tau^-$ background events than 
$t\bar{t}$ signal
events.
Therefore, we developed a selection that targets this
background exclusively.  $Z\to\tau^+\tau^-$ decays that pass the $H_T$ and
$\met$ requirements have a $Z$ boson with significant 
$p_T$.
In these
events, one can reconstruct the $Z$ mass from the observed tau decay
products by assuming that the $\met$ in the events results entirely
from neutrinos produced in $\tau$ decays and that those neutrinos are
collinear with the other $\tau$ decay products.  
 In this case, there is a
unique assignment of the energy of unobserved neutrinos from each tau
candidate based on the direction of the $\met$.  
{\ifnum\cutText=0
 For there to be a sensible solution, the $\met$ must be able to be
 decomposed into components parallel to the direction of both
the electron or muon and the tau candidate, i.e. the $\met$ must
point in a direction between the transverse momenta of the electron or
muon and the tau candidate.
\fi} 
For $t\bar{t}$ events, this is most often not
the case, and the assignment of $\met$ to a neutrino would result in
the neutrino carrying negative momentum from the tau decay.
For events where this reconstruction is sensible,
we remove events in a window 25 GeV above and below the $Z$ mass.  This results in an
additional
$88\%$ reduction of $Z\to\tau^+\tau^-$ events while removing
only $4\%$ of
$t\bar{t}$ signal.
{\ifnum\cutText=0
\\
The other physics processes that can reproduce 
the
 final state with
taus are $WW$ and $WZ$ 
bosons
created in association with jets.  The
product of the theoretical cross-section and branching ratios for these
processes is much smaller than that of $t\bar{t}$, and they are also
removed preferentially by the $H_T$ requirement.  
We predict the
background based on a simulation of diboson production from the HERWIG
generator~\cite{herwig}.  An additional source of $t\to\tau \nu q$ 
signal events in electroweak single top production is
expected to contribute fewer than $0.01$ events.
\fi} 
We summarize all backgrounds in Table~\ref{tab:results1}.

\begin{table}[tbp]
\begin{center}
{\ifnum\cutText=0
{\small
\begin{tabular}{lc}
\hline
\hline
Process & Number of expected events\\
\hline
$\gamma^*/Z \rightarrow \tau\tau +$ jets & 0.25 $\pm$0.06 $\pm$0.05 \\
W/Z + jets with jet $\rightarrow\tau$ fake & 0.75 $\pm$0.12 $\pm$0.20\\
$\gamma^*/Z\rightarrow ee$ + jets with $e \rightarrow \tau$ fake & 0.08 $\pm$0.03 $\pm$0.02\\
$\gamma^*/Z\rightarrow \mu\mu$ + jets with $\mu\rightarrow\tau$ fake  & 0.05 $\pm$0.03$\pm$0.01 \\
$WW$ & 0.14 $\pm$0.02 $\pm$0.03 \\
$WZ$ & 0.02 $\pm$0.02$\pm$0.01 \\
\hline 
Total expected background events &   1.29 $\pm$0.14 $\pm$ 0.21\\
\hline
Expected signal & 1.00 $\pm$0.06 $\pm$0.16\\
\hline
\hline
\end{tabular}}
\else
{\small
\begin{tabular}{lc}
\hline
\hline
Process & Number of expected events\\
\hline
$\gamma^*/Z \rightarrow \tau\tau +$ jets & 0.25 $\pm$0.06 $\pm$0.05 \\
W/Z + jets with jet $\rightarrow\tau$ fake & 0.75 $\pm$0.12 $\pm$0.20\\
$\gamma^*/Z\rightarrow ee$ + jets with $e \rightarrow \tau$ fake & 0.08 $\pm$0.03 $\pm$0.02\\
$\gamma^*/Z\rightarrow \mu\mu$ + jets with $\mu\rightarrow\tau$ fake  & 0.05 $\pm$0.03$\pm$0.01 \\
$WW$ & 0.14 $\pm$0.02 $\pm$0.03 \\
$WZ$ & 0.02 $\pm$0.02$\pm$0.01 \\
Single top, $t\to\tau\nu q$ & $<$ 0.01 \\
\hline 
Total expected background events &   1.29 $\pm$0.14 $\pm$ 0.21\\
\hline
Expected signal & 1.00 $\pm$0.06 $\pm$0.16\\
\hline
\hline
\end{tabular}}
\fi}
\end{center}
\caption{Summary of background and signal predictions.  The first
error is from simulation and data statistics, and the second is from
systematic uncertainties.  
{\ifnum\cutText>0
Diboson backgrounds are predicted in a
simulation based on the HERWIG generator~\cite{herwig}.  
\fi}
The expected signal assumes a $t\bar{t}$ 
cross-section of $6.7$~pb.}
\label{tab:results1}
\end{table}

We use an independent control sample 
to test our calculation of backgrounds.  The selection of the
electrons, muons, taus and $\met$ is identical to our signal
selection, but the number of jets in the sample is restricted to be
less than two.  Also, in order to increase the statistics for this comparison
we do not impose the $H_T$ 
requirement or the $Z$ mass removal on these events.
Table~\ref{tab:njets} shows the comparisons of predicted and observed
events.  We categorize the results based on jet multiplicity, electron
or muon final state, and the cases of the same or opposite charge in
the two leptons.

The data in Table~\ref{tab:njets} can be used as a control experiment
to check the accuracy of our background predictions.  {\em A priori},
we chose as the statistic of the control experiment the
joint probability 
of the observed number of events given the
prediction for
the
eight samples in this table.  The expected
distribution of these joint probabilities is measured 
via
simulated pseudo-experiments which account for the uncertainties in the
predictions and the Poisson fluctuations from the limited
statistics in the data.  We 
find
the data in Table~\ref{tab:njets} 
have
a joint probability which 
is
higher than 41\% of our
pseudo-experiments and conclude that these control data are consistent with our
expected background predictions.

\begin{table}
{\ifnum\cutText=0
\begin{tabular}{lccc}
\hline
\hline
channel &~~ \# jets &~~ predicted \# events &~~ measured \# events\\ \hline
e + $\tau$ opp sign & 0 & 23.7$\pm$3.6  & 17 \\
e + $\tau$ opp sign & 1 & 4.6$\pm$0.9 & 5 \\
e + $\tau$ same sign & 0 & 7.3$\pm$1.8 & 8 \\
e + $\tau$ same sign & 1 & 1.9$\pm$0.6  & 3 \\
$\mu$ + $\tau$ opp sign & 0 & 21.3$\pm$3.3 & 11 \\
$\mu$ + $\tau$ opp sign & 1 & 2.7$\pm$0.6 & 4 \\
$\mu$ + $\tau$ same sign & 0 & 5.6$\pm$1.5 & 3  \\
$\mu$ + $\tau$ same sign & 1 & 0.8$\pm$0.3 & 0 \\
\hline
\hline
\end{tabular}
\else
\begin{tabular}{l|cc|cc}
\hline
\hline
& \multicolumn{2}{c} events predicted & \multicolumn{2}{|c} events measured \\
channel & 0 jets & 1 jet & 0 jets & 1 jet \\ \hline
e + $\tau$ opp sign & 23.7$\pm$3.6 & 4.6$\pm$0.9  & 17 & 5 \\
e + $\tau$ same sign & 7.3$\pm$1.8 & 1.9$\pm$0.6  & 8 & 3 \\
$\mu$ + $\tau$ opp sign & 21.3$\pm$3.3 & 2.7$\pm$0.6 & 11 & 4\\
$\mu$ + $\tau$ same sign & 5.6$\pm$1.5 & 0.8$\pm$0.3 & 3 & 0  \\
\hline
\hline
\end{tabular}
\fi} 
\caption{Comparison between the predicted and measured number of
events for low jet multiplicities}
\label{tab:njets}
\end{table}

\begin{table}
{\ifnum\cutText=0
\begin{tabular}{lcc}
\hline
\hline
&~~ Candidate \#1 &~~ Candidate \#2 \\ \hline
tau $p_T$ (GeV/$c$) & 39 & 20 \\
electron $p_T$ (GeV/$c$) & 40 & 79 \\
\# of tau tracks & 1  & 3 \\
$H_T$ (GeV) & 286 & 239 \\
Number of jets & 3 & 2 \\
jet1 $E_T$ (GeV) & 73 $\dagger$ & 35 \\
jet2 $E_T$ (GeV) & 40 & 34 \\
jet3 $E_T$ (GeV) & 35 & n/a \\
$\met$ (GeV) & 59  & 72 \\
\hline
\hline
\end{tabular}
\else
\begin{tabular}{lcc}
\hline
\hline
&~~ Candidate \#1 &~~ Candidate \#2 \\ \hline
tau $p_T$ (GeV/$c$) & 39 & 20 \\
electron $p_T$ (GeV/$c$) & 40 & 79 \\
\# of tau tracks & 1  & 3 \\
$H_T$ (GeV) & 286 & 239 \\
$E_T$ of jets (GeV) & 73$\dagger$, 40, 35 & 35, 34 \\
$\met$ (GeV) & 59  & 72 \\
\hline
\hline
\end{tabular}
\fi} 
\caption{Properties of the candidate events.  The jet marked with the
  $\dagger$ is tagged as a $b$ by the presence of a secondary
  vertex~\cite{ttbar-secvtx-prd}.} 
\label{tab:events}
\end{table}

{\ifnum\cutText=0
The theoretical $t\bar{t}$ production cross-section is
 $6.7^{+0.7}_{-0.9}$~pb~\cite{ttbar-xsect,ttbar-xsect2} for a top quark mass
 of $175$~GeV/$c^2$.
\fi}
The signal acceptance,  including all branching fractions
in the decay channel, 
is
$0.076\pm0.005{\rm\textstyle (stat.)}
\pm0.013{\rm\textstyle (syst.)}\%$.  
For a $t\bar{t}$ cross-section of $6.7$~pb~ we therefore expect $1.00\pm0.17$
signal events in $194$~pb$^{-1}$, in addition to the background
expectation of $1.29\pm0.25$ events. We observe 2 events. 
Both events are in the 
electron + tau channel, and properties of these two events are listed in
Table~\ref{tab:events}.

We define the ratio of partial widths ($\Gamma$)
$$
r_\tau\equiv  \frac{{\Gamma}(t\to \tau\nu q)}{{\Gamma_{SM}}(t\to
\tau\nu q)}
$$
as a measure of a possible anomalous enhancement in the rate.  Using
the method of Rolke 
%{\em et alia}~\cite{Rolke}, 
{\em et al.}~\cite{Rolke},
we set an upper limit 
on
$r_\tau$ of $5.2$ at 
the 95\% confidence level.
The previous result from CDF Run I~\cite{r1tau} would yield
a limit of $r_\tau<18$ at 
the 95\% confidence level using the same statistical methods.

In summary, we have searched for top decay into $\tau\nu q$ by
identifying semi-hadronic decays of tau leptons in $t\bar{t}$ events.  
We observe two candidate
events, consistent with 
the standard model, 
and set an upper limit on the 
ratio of observed production of  $t\to\tau\nu q$
to the standard model expectation.

\begin{acknowledgments}
We thank the Fermilab staff and the technical staffs of the
participating institutions for their vital contributions. This work
was supported by the U.S. Department of Energy and National Science
Foundation; the Italian Istituto Nazionale di Fisica Nucleare; the
Ministry of Education, Culture, Sports, Science and Technology of
Japan; the Natural Sciences and Engineering Research Council of
Canada; the National Science Council of the Republic of China; the
Swiss National Science Foundation; the A.P. Sloan Foundation; 
the Research Corporation;
the Bundesministerium f\"ur Bildung und Forschung, Germany; the Korean
Science and Engineering Foundation and the Korean Research Foundation;
the Particle Physics and Astronomy Research Council and the Royal
Society, UK; the Russian Foundation for Basic Research; the Comisi\'on
Interministerial de Ciencia y Tecnolog\'{\i}a, Spain; in part by the
European Community's Human Potential Programme under contract
HPRN-CT-2002-00292; and the Academy of Finland.

\end{acknowledgments}


\begin{thebibliography}{99}

\bibitem{PDG} Particle Data Group, \pl{B592}{1}{2004}
\bibitem{higgs} V. Barger and R.J.N. Phillips, \prev{D41}{884}{1990}
\bibitem{susyHiggs-top} J. Guasch and J. Sola, \pl{B416}{353}{1998}
\bibitem{higgs2} T. Affolder {\em et al.\ }[CDF collaboration],
  \prev{D62}{012004}{2000}
\bibitem{rpv-top} T. Han and M.B. Magro, \pl{B476}{79}{2000}
\bibitem{nu-zprime} C. Yue, H. Zong and L. Liu, Mod.\ Phys.\ Lett.\ {\bf
  A18}, 2187 (2003).
\bibitem{cdf-top-disc}
F.~Abe {\it et al.}  [CDF collaboration],
%``Observation of top quark production in anti-p p collisions,''
Phys.\ Rev.\ Lett.\  {\bf 74}, 2626 (1995).
\bibitem{d0-top-disc}
S.~Abachi {\it et al.}  [\D0  collaboration],
%``Observation of the top quark,''
Phys.\ Rev.\ Lett.\  {\bf 74}, 2632 (1995).
%\bibitem{herwig} HERWIG 6.5, G. Corcella, I.G. Knowles, G. Marchesini, S. Moretti, K. Odagiri, P. Richardson, M.H. Seymour and B.R. Webber, JHEP 0101 (2001) 010 [hep-ph/0011363]; hep-ph/0210213.
\bibitem{ttbar-xsect}
M.~Cacciari, S.~Frixione, M.~L.~Mangano, P.~Nason and G.~Ridolfi,
%``The t anti-t cross-section at 1.8-TeV and 1.96-TeV: A study of the
%systematics due to parton densities and scale dependence,''
JHEP {\bf 0404}, 068 (2004).
\bibitem{ttbar-xsect2}
N.~Kidonakis and R.~Vogt,
Phys.\ Rev.\ {\bf D68}, 114014 (2003).
\bibitem{r1tau} F. Abe {\em et al.\ }[CDF collaboration],\prl{79}{3585}{1997}
\bibitem{var} CDF uses a cylindrical coordinate system in which $\phi$ is the
azimuthal angle, $r$ is the radius from the nominal beamline, and $+z$
points in the direction of the proton beam and is zero at the center of the
detector.  The pseudorapidity $\eta = -ln[\tan(\theta / 2)]$, where $\theta$
is the polar angle with respect to the $z$ axis.  Calorimeter energy
(track momentum) measured transverse to the beam is denoted as $E_T$ 
($p_T$). 
\bibitem{det}  
D. Acosta, {\em et al.\,}[CDF collaboration], Phys.\ Rev.\ {\bf D71}, 032001 (2005). 
%D. Acosta {\em et al.\,}[CDF collaboration], 
%Fermilab-Pub-04-440-E., arXiv:hep-ex/0412071, submitted to Phys.\
%Rev.\ {\bf D}.
\bibitem{ttbar-kinem-prd} D. Acosta {\em et al.\ }[CDF collaboration],
% ``Measurement of the t anti-t production cross section in p anti-p collisions
%at s**(1/2) = 1.96-TeV using kinematic fitting of b-tagged lepton + jet
%events,''
  Phys.\ Rev.\ {\bf D71}, 072005 (2005).
\bibitem{wz-prod-prl} D. Acosta {\em et al.\ }[CDF collaboration],
  %``First measurements of inclusive W and Z cross sections from Run II of the
%Tevatron collider,''
%arXiv:hep-ex/0406078, submitted to Phys.\ Rev.\ Lett.
\prl{94}{091803}{2005}
\bibitem{dilep-prl} D. Acosta {\em et al.\ }[CDF collaboration],
  \prl{93}{142001}{2004}
\bibitem{pythia} T. Sjostrand, Comput.\ Phys.\ Commun. {\bf 82}, 74 (1994).
\bibitem{tauola} S. Jadach {\em et al.\,}, CERN-TH-6793 (1992).
\bibitem{GEANT4} S.~Agostinelli {\em et al.\,} [GEANT4
  collaboration], Nucl.\ Inst.\ Meth.\ {\bf A506} 250 (2003).
\bibitem{alpgen} M. Mangano, {\it et al.}, JHEP {\bf 07 }, 001 (2003).
\bibitem{herwig} G. Corcella, {\it et al.}, JHEP {\bf 0101}, 010 (2001).
\bibitem{ttbar-secvtx-prd} D. Acosta {\em et al.\ }[CDF
  collaboration],
%FERMILAB-PUB-04-275-E, arXiv:hep-ex/0410041, submitted to Phys.\ Rev.\ {\bf D}.
 Phys.\ Rev.\ {\bf D71}, 052003 (2005).
\bibitem{Rolke} 
W.~A.~Rolke, A.~M.~Lopez, and J.~Conrad,
%``Confidence Intervals with Frequentist Treatment of Statistical and
%Systematic Uncertainties,''
 Nucl.\ Inst.\ Meth.\ {\bf A539} 407 (2005).
%arXiv:physics/0403059.

\end{thebibliography}
\end{document}